\documentclass[11pt,preprintnumbers,amsmath,amssymb]{revtex4}
\usepackage{amssymb}
\usepackage{amsfonts}
\usepackage{mathrsfs}
\usepackage{amsmath}
\usepackage{graphicx}
\usepackage{dcolumn}
\usepackage{bm}
\usepackage{hyperref}

\begin{document}
\title{Hacking on decoy-state quantum key distribution system with partial phase randomization}
\author{Shi-Hai Sun \footnote{email:shsun@nudt.edu.cn}, Mu-Sheng Jiang, Xiang-Chun Ma, Chun-Yan Li, Lin-Mei Liang}
\affiliation{Department of Physics, National University of Defense
Technology, Changsha 410073, P.R.China}
\begin{abstract}
Quantum key distribution (QKD) provides means for unconditional secure key transmission between two distant parties. However, in practical implementations, it suffers from quantum hacking due to device imperfections. Here we propose a hybrid measurement attack, with only linear optics, homodyne detection, and single photon detection, to the widely used vacuum+weak decoy state QKD system when the phase of source is partially randomized. Our analysis shows that, in some parameter regimes, the proposed attack would result in an entanglement breaking channel but still be able to trick the legitimate users to believe they have transmitted secure keys. That is, the eavesdropper is able to steal all the key information without discovered by the users. Thus, our proposal reveals that partial phase randomization is not sufficient to guarantee the security of phase-encoding QKD systems with weak coherent states.
\end{abstract}

\newpage
\maketitle
Quantum key distribution (QKD) \cite{BB84} admits two remote parties (Alice and Bob) to share unconditional secure key based on the principle of quantum mechanics \cite{Shor00,GLLP04}, which has been demonstrated in experiments with long distance and high repetition rate \cite{Wang12,Namekata09,Yuan08,Liu10}. However, the practical QKD system will suffer from quantum hacking due to device imperfections \cite{Sun11,Sun12,Gerhardt11,Lydersen10,Jain11,Xu10,Qi07,Tang13}, then the unconditional security of QKD is compromised. In practical QKD systems based on BB84 protocol, the weak coherent source (WCS) is often used to replace the single photon source which is unavailable within current technology. However, the WCS contains multi-photon pulse with nonzero probability which will cause the photon-number-splitting (PNS) attack \cite{Huttner95,Brassard00}, then the maximal secure distance of practical QKD system will be limited in tens of kilometers. Luckily, decoy state method \cite{Hwang03,Wang05,Lo05,Ma05} can efficiently overcome this problem, and extend the secure distance of QKD to hundreds of kilometers.

When the phase of WCS has been totally randomized, the source is a mixed state of all number states, and the channel between Alice and Bob can be considered as a photon number channel. Then, the key rate is given by the GLLP formula \cite{GLLP04},
\begin{equation}\label{GLLP}
R=q\{-Q_\mu f(E_\mu) H_2(E_\mu)+\mu e^{-\mu}Y_1^L[1-H_2(e_1^U)]\},
\end{equation}
where $q=1/2$ for the standard BB84 protocol, $H_2(x)$ is binary Shannon entropy, $f(E_{\mu})$ is the error correction efficiency. $Q_{\mu}$ and $E_\mu$ are the total gain and QBER, which can be measured in experiment. $Y_1^L$ and $e_1^U$ are the lower bound of yield and upper bound of QBER for single photon pulses, which must be estimated by Alice and Bob according to their measurement results. In fact, the main contribution of decoy state method is that it can give out the tight bound of $Y_1$ and $e_1$ with finite resources. For instance, the weak+vacuum decoy state method is enough for the legitimate parties to tightly estimate the yield and QBER of single photon pulses, in which Alice randomly sends three kinds of pulses with different intensities, signal state $\mu$, decoy state $\nu$, and vacuum state. After the communication, Alice and Bob calculate the total gain ($Q_\mu$, $Q_\nu$ and $Q_{vac}$) and QBER ($E_\mu$, $E_\nu$ and $E_{vac}$) in experiment, then they estimate the lower bound of yield ($Y_1^L$) and the upper bound of QBER ($e_1^U$) for the single photon pulse, which are given by \cite{Ma05}
\begin{equation}
\begin{split}
Y_1^L&=\frac{\mu}{\mu\nu-\nu^2}(Q_\nu e^{\nu}-Q_\mu e^\mu \frac{\nu^2}{\mu^2}-\frac{\mu^2-\nu^2}{\mu^2}Q_{vac}),\\
e_1^U&=\frac{E_\nu Q_\nu e^{\nu}-E_{vac}Q_{vac}}{Y_1^L \nu}.
\end{split}
\end{equation}

Obviously, the phase randomization is the base of decoy state method. However, in practical situations, this assumption may not hold, since Eve may have some prior information about the random phase of source. For example, in two-way systems, the source is totally controlled by Eve, thus she can exactly know the phase of source; or in some systems, the pulse is generated by cutting off the coherent laser with a intensity modulation, and there may exits phase relationship among different pulses. In fact, some potential attack on source had been proposed \cite{Sun12,Tang13,Lo07}. In Ref.\cite{Lo07}, Lo and Preskill pointed out that the phase randomization assumption is necessary for the security of BB84 protocol using WCS, and obtained the key rate formula with nonrandom phase. In Ref.\cite{Tang13}, Tang \emph{et al.} proposed and demonstrated an attack, based on a linear-optic unambiguous state discrimination measurement and PNS, to show that the security of a QKD system with nonrandom phase will be compromised. In Ref.\cite{Sun12}, our group proposed an attack to show that the QKD system is still insecure even if the phase of source is partially randomized, but it is invalid for the widely used weak+vacuum decoy state method (their attack is only valid for the special one-decoy state method in some parameter regimes).

In this paper we propose a more powerful hybrid measurement attack, with only linear optics, homodyne detection, and single photon detection (SPD), to the widely used vacuum+weak decoy state QKD system when the phase of source is partially randomized. Here partial phase randomization means that the phase of source is randomized within the range of $[0,\delta)$, where $\delta\leq2\pi$. Note that $\delta=0$, $\delta<2\pi$ and $\delta=2\pi$ represents unrandomization, partial randomization and total randomization, respectively. When the phase of source is just partially randomized, the photon number channel assumption, which is the base of the decoy state, is invalid, then Eve can use this information to enhance her ability to spy the secret key. Our analysis shows that the proposed attack would result in an entanglement breaking channel but still be able to trick the legitimate users to believe they have transmitted secure keys. That is, the eavesdropper is able to steal all the key information without noticed by the users. Thus, our proposal reveals that partial phase randomization is not sufficient to guarantee the security of phase-encoding QKD systems with coherent states.

Furthermore, we remark that, recently, the measurement device independent (MDI-) QKD is proposed \cite{Lo13} and demonstrated \cite{Liu13,Robenok13} to exclude all the detection loopholes, but it requires that the source can be fully characterized. Specially, when WCS is used in practical MDI-QKD sytems, it also needs to ensure that the phase of source is totally randomized, otherwise, the decoy state method (weak+vacuum decoy state method) \cite{Wang13,Ma12,Xu13,Sun13} can not be applied to estimate the key rate. Thus we think that our work is also significant for the MDI-QKD.

\textbf{Results}
\begin{figure}
\scalebox{1}{\includegraphics[width=8cm]{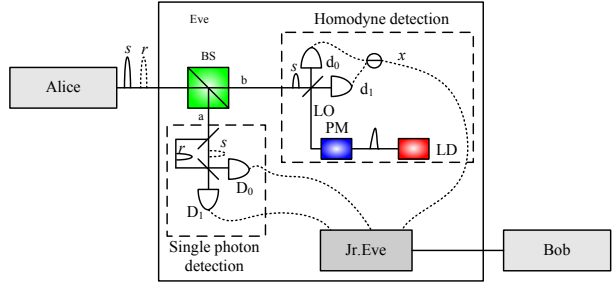}}
\caption{\label{fig:scheme}The diagram of the hybrid measurement attack. $r$($s$) is the signal (reference) pulse of Alice. BS: beam splitter with transmittance 1/2; $D_0$ and $D_1$ are single photon detectors (SPDs); $d_0$ and $d_1$ are photodiodes; $x$ is the output of homodyne detection; LD: laser diode which is used by Eve to generate the reference pulse (LO pulse) of homodyne detection; PM: phase modulator which is used by Eve to modulate a phase (0 or $\pi/2$) on LO. Jr.Eve has the same equipments as Alice, which is used to resend faked states to Bob according to her measurement results. Note that, Eve measures both $r$ and $s$ of Alice with a interferometer in the single photon detection part, but she only measures the phase information of $s$ in the homodyne detection part.}
\end{figure}

A diagram of our hybrid measurement attack is shown in Fig.\ref{fig:scheme}. Eve first splits Alice's pulses (both $r$ and $s$) into two parts with a beam splitter (BS). Without loss generality, here we assume the transmittance of BS is 1/2, and label the reflected part as $a$ and transmitted part as $b$. For the part $a$, Eve lets $r$ and $s$ to interfere with an asymmetry interferometer, then she records the results with two single photon detectors ($D_0$ and $D_1$). For the part $b$, Eve generates a strong reference pulse (LO pulse) with her own laser diode (LD), and randomly modulates a phase ($\phi_e=0,\pi/2$) on the LO pulse with a phase modulator (PM). Then she lets $s$ to interfere with the LO pulse, and records the results with a homodyne detection which is composed with two photodiodes ($d_0$ and $d_1$) and a subtracter. Note that, $r$ is neglected in homodyne detection part, since it does not carry the encoding phase of Alice. Furthermore, excepting phase information, the LO pulse generated by Eve should be indistinguishable with the $s$ in frequency, polarization and other dimensions. We think it is possible for Eve to generate the indistinguishable pulse with Alice, since, excepting phase information, other characters of Alice's laser are excluded in the secure model of Alice and can be known by Eve.

Now we give an explanation of our attack and show that it can be applied to the widely used weak+vacuum decoy state method. In BB84 protocol with WCS, the state of Alice can be written as $|\alpha e^{i(\theta+\phi)}/\sqrt{2}\rangle_s|\alpha e^{i\phi}/\sqrt{2}\rangle_r$, where $\alpha$ is real and $|\alpha|^2=\mu$ is the intensity of Alice's pulse, $\theta=\{0,\pi/2,\pi,3\pi/2\}$ is the encoding phase of Alice, $\phi\in [0,\delta)$ is the random phase of source and $\delta$ is the range of phase randomization. According to the measurement theory, the probability that $D_0$ and $D_1$ click in the single photon detection part and measurement result $x$ is obtained in the homodyne detection part are given by
\begin{equation}\label{detec}
\begin{split}
P_{D_0}&=1-(1-Y_0^E)e^{-\mu\eta_E[1+cos(\theta)]/4},\\
P_{D_1}&=1-(1-Y_0^E)e^{-\mu\eta_E[1-cos(\theta)]/4},\\
P_x(\theta,\phi,\phi_e)&=\sqrt{\frac{2}{\pi\kappa_E^2}}e^{-2[x-\lambda_E|\alpha|cos(\theta+\phi-\phi_e)/2]^2/\kappa_E^2},
\end{split}
\end{equation}
where $Y_0^E$ ($\eta_E$) is the dark count (detection efficient) of Eve's SPDs,  $\phi_e=0,\pi/2$ is the phase modulated by Eve on the LO pulse with PM, $\kappa_E$ and $\lambda_E$ represent the imperfection of Eve's homodyne detection ($\kappa_E=\lambda_E=1$ for perfect homodyne detection).

According to Eq.\ref{detec}, $P_{D_0}$ and $P_{D_1}$ are independent on the random phase $\phi$, but $P_x(\theta,\phi,\phi_e)$ depends on $\phi$. Since Eve has no prior information about $\phi$ excepting that $\phi\in[0,\delta)$, thus the probability distribution of $x$ should be written as
\begin{equation}\label{Px}
P_x(\theta,\phi_e)=\int_0^\delta \frac{d\phi}{\delta}P_x(\theta,\phi,\phi_e).
\end{equation}

\begin{figure}
\scalebox{1}{\includegraphics[width=8cm]{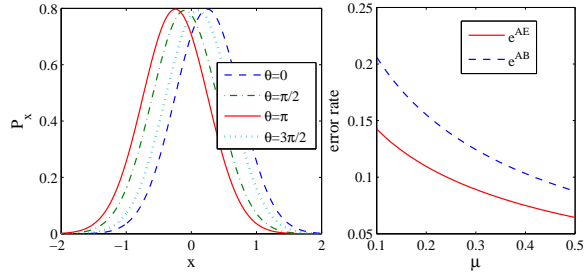}}
\caption{\label{fig:x}(a)The theoretical distribution of $x$ for different encoding phase of Alice, which are drawn according to Eq.\ref{Px}. Here we assume $\phi_e=0$, $\delta=\pi/4$ and $\mu=0.3$. (b) The error rate of Eve and Bob under our attack, which are drawn according to Eq.\ref{PE}. The solid line shows the error rate between Alice and Eve, and the dashed line shows the error rate between Alice and Bob. Here we set $\delta=10^\circ$, $x_0=1.5$, and assume that the detection setups of both Alice and Bob are perfect.}
\end{figure}

The theoretical distribution of $x$ is shown in Fig.\ref{fig:x}(a), which clearly shows that Eve can use $x$ to distinguish encoding phase of Alice. For example, Eve can set a threshold ($x_0>0$), when the measured $x$ is larger than $x_{0}$, she judges that $\theta=0$, and when $x<-x_0$, she judges that $\theta=\pi$, otherwise ($-x_0<x<x_0$), she randomly guess Alice's bit. Note that, in BB84 protocol, Alice randomly chooses her phase from two bases, thus Eve also should randomly modulate a phase ($\phi_e=0,\pi/2$) on the LO pulse with a PM to judge which basis is used by Alice. In fact, this part is the same as the partially random phase (PRP) attack proposed by our group \cite{Sun12}, however, the PRP attack is invalid for the weak+vacuum decoy state method due to the fact that the homodyne detection will export a successful result ($x>x_0$ or $x<-x_0$) with high probability, even if a vacuum state is sent by Alice, thus the total gain and QBER are much larger than the expectation of Bob without Eve. In order to reduce the disadvantage of homodyne detection, we introduce an additional measurement for Eve. Eve uses an interferometer and two SPDs to judge whether there is photon in Alice's pulse or not. Only when one of her SPD clicks, she resends a faked state to Bob, otherwise, she resends a vacuum state to Bob. Therefore, the mapping from Eve's measurement results to the phase of her faked state ($\theta_e$) is given by
\begin{equation}
\begin{split}
\phi_e=0&
\begin{cases}\text{$x>x_0$ and $P_{D_0}$ click} &\rightarrow \theta_e=0,\\
 \text{$x<-x_0$ and $P_{D_1}$ click}& \rightarrow \theta_e=\pi,\\
 \text{otherwise} & \rightarrow\text{vacuum pulse}.
 \end{cases}\\
 \phi_e=\pi/2&
\begin{cases}\text{$x>x_0$ and $P_{D_0}$ click} &\rightarrow \theta_e=\pi/2,\\
 \text{$x<-x_0$ and $P_{D_1}$ click}& \rightarrow \theta_e=3\pi/2,\\
 \text{otherwise} & \rightarrow\text{vacuum pulse}.
 \end{cases}
 \end{split}
\end{equation}
And the conditional probability that Eve resends the state with phase $\theta_e=k\pi/2$ ($k=0,1,2,3$) given that Alice sends a state with phase $\theta$ is given by
\begin{equation}
\begin{split}
P_e^{0|\theta}&=\frac{1}{2}P_{D_0}\int_{x_0}^\infty dx \int_0^\delta \frac{d \phi}{\delta}P_x(\theta,\phi,\phi_e=0),\\
P_e^{\pi/2|\theta}&=\frac{1}{2}P_{D_0}\int_{x_0}^\infty dx \int_0^\delta \frac{d \phi}{\delta}P_x(\theta,\phi,\phi_e=\pi/2),\\
P_e^{\pi|\theta}&=\frac{1}{2}P_{D_1}\int_{-\infty}^{-x_0} dx \int_0^\delta \frac{d \phi}{\delta}P_x(\theta,\phi,\phi_e=0),\\
P_e^{3\pi/2|\theta}&=\frac{1}{2}P_{D_1}\int_{-\infty}^{-x_0} dx \int_0^\delta \frac{d \phi}{\delta}P_x(\theta,\phi,\phi_e=\pi/2).
\end{split}
\end{equation}
Thus, when Eve is present, the probability that she successfully obtains a measurement event, the QBER between Alice and Bob ($e^{AB}$), and the QBER between Alice and Eve ($e^{AE}$) are given by
\begin{equation}\label{PE}
\begin{split}
P_{succ}^E&=\frac{1}{4}\sum_{j=0}^3\sum_{k=0}^3 P_e^{\frac{k\pi}{2}|\frac{j\pi}{2}},\\
e^{AB}&=\frac{1}{4}\sum_{j=0}^3 \frac{\sum_{k=0}^3 P_e^{\frac{k\pi}{2}|\frac{j\pi}{2}}e^{AB}_{k|j}}{\sum_{k=0}^3 P_e^{\frac{k\pi}{2}|\frac{j\pi}{2}}},\\
e^{AE}&=\frac{1}{4}\sum_{j=0}^3 \frac{\sum_{k=0}^3 P_e^{\frac{k\pi}{2}|\frac{j\pi}{2}}e^{AE}_{k|j}}{\sum_{k=0}^3 P_e^{\frac{k\pi}{2}|\frac{j\pi}{2}}},
\end{split}
\end{equation}
where $e^{AB}_{k|j}$ is the error rate introduced by Eve's faked state with phase $j\pi/2$ given that Alice's phase is $\theta=j\pi/2$. $e^{AE}_{k|j}$ is the error rate of Eve for given $k$ and $j$. The error rate $e^{AB}$ and $e^{AE}$ are shown in Fig.\ref{fig:x}(b), which clearly shows that the error rate between Alice and Eve is much smaller than the error rate between Alice and Bob. Here we remark that although $e^{AE}$ is smaller than $e^{AB}$, it does not means no secret key can be derived due to the fact that post-processing is not symmetric between Eve and Bob. In fact, if we want to show our attack is succeed and the QKD system is insecure, we must show that the lower bound of the estimated key rate given that Eve implements her attack but the legitimate parties ignore it is larger than the upper bound of key rate under the given attack \cite{Tang13}. For example, our analysis shows that, when our attack is implemented but the legitimate parties ignore it, the estimated key rate per pulse by Alice and Bob can be larger than $10^{-3}$ in some parameters regimes, but in fact our attack belongs to intercept-and-resend attack (Eve measures all the signals and resend her prepared pulses to Bob), which corresponds to an entanglement-breaking channel and no secret key can be generated under this channel. In other words, the upper bound of key rate under our attack is zero. Thus all the estimated key are insecure. In the following, we give a detailed analysis.

Since Eve can not distinguish the signal state, decoy state and vacuum state, thus we assume that Eve resends a single photon state to Bob when she successfully obtains a measurement event. In other words, the total gain and QBER under our attack are given by
\begin{equation}\label{QE}
\begin{split}
Q_\omega&= \eta_{Bob}P_{succ}^E +(1-P_{succ}^E\eta_{Bob})Y_0,\\
Q_\omega E_\omega&=\eta_{Bob}P_{succ}^Ee_{Eve} +(1-P_{succ}^E\eta_{Bob})Y_0e_0.
\end{split}
\end{equation}
where $\omega=\{\mu,\nu,0\}$, $Y_0$ is the dark count of Bob's SPD, $e_0=1/2$ is the error rate of background, and $\eta_{Bob}$ is the transmittance of Bob's setups. $P_{succ}^E$ and $e_{Eve}=e^{AB}$ are given by Eq.\ref{PE} for different intensity of pulses.

\begin{figure}
\scalebox{1}{\includegraphics[width=8cm]{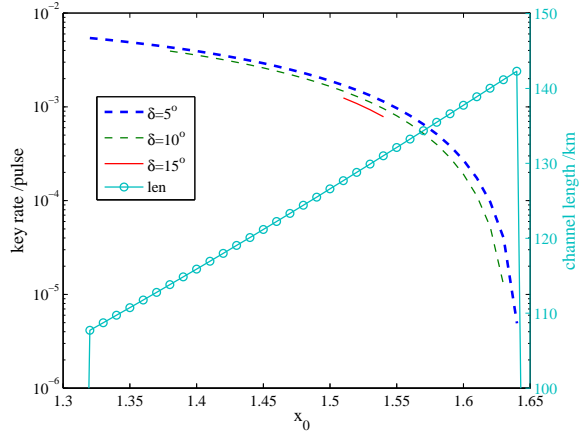}}
\caption{\label{fig:key_rate}The estimated key rate of Alice and Bob under our attack. But in fact, the key are insecure, since our attack corresponds to an entanglement-breaking channel and no secret key can be generated under this channel.  Here we also show the equivalent channel length of $Q_\mu$, defined as $len=-(10/a)\log_{10}\{\min(1,Q_\mu/(\mu \eta_{Bob})\}$ ($a=0.21$ is the loss of standard fiber), which represents the minimal channel length of Alice and Bob that Eve can successfully load our attack. In the simulations, we assume that the SPD and homodyne detection of Eve are perfect, and set $f(E_\mu)=1.22$, $Y_0=1.7\times10^{-6}$, $\eta_{Bob}=0.045$, $\mu=0.48$, and $\nu=0.1$ according to the experimental results of Ref. \cite{Yuan08}.}
\end{figure}

\begin{figure}
\scalebox{1}{\includegraphics[width=8cm]{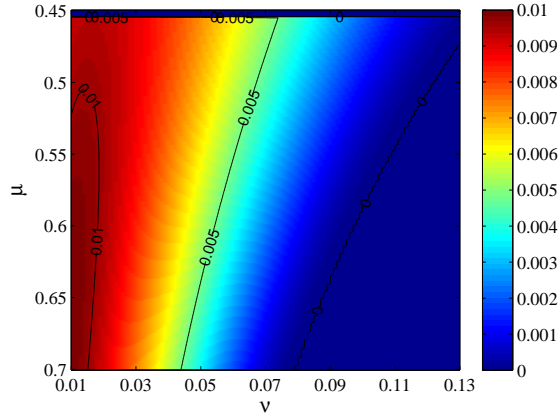}}
\caption{\label{fig:key_rate_uv}The estimated key rate of Alice and Bob for different $\mu$ and $\nu$ when Eve is present. In the simulations, we set $x_0=1.5$, $\delta=10^\circ$, and other parameters are the same as Fig.\ref{fig:key_rate}.}
\end{figure}

By substituting Eq.\ref{QE} into Eq.\ref{GLLP}, we can estimate the key rate under our attack, which is shown in Fig.\ref{fig:key_rate}. It clearly shows that even Eve is present, Alice and Bob still can obtain positive key rate. For example, when $\delta=10^\circ$, the key rate is positive if Eve sets $1.38<x_0<1.63$. However, these key are insecure in this range, since our attack corresponds to an entanglement-breaking channel and no secret key can be generated under this channel. Furthermore, we estimate the key rate for different intensities of signal state and decoy state in Fig.\ref{fig:key_rate_uv}, which also clearly shows that our attack is valid in some parameter regimes.

\textbf{Discussion}

According to the analysis above, we know that when the phase of source is partially randomized, the security of the widely used weak+vacuum decoy state QKD will be compromised. Our attack shows that, in some parameter regimes, when Eve is present, the legitimate parties will be cheated and the estimated key rate is still positive, but in fact, the generated key are insecure, since our attack belongs to intercept-and-resend attack (Eve measures all the signals and resend her prepared pulses to Bob), which corresponds to an entanglement-breaking channel and no secret key can be generated under this channel. Here we remark that, we do not claim our attack is optimal for Eve to exploit the partially random phase of source, in fact our attack is valid just in some given parameter regimes. However, our attack still plays an important role in reminding the legitimate users that, phase randomization is necessary to guarantee the security of practical QKD system with WCS, and, instead of calibrating the random phase before the communication, they must carefully consider the phase randomization assumption and ensure that this assumption hold in the communication progress, otherwise their system may be insecure.

In the end we discuss three countermeasures. The first one is that Alice uses an active phase randomization equipment \cite{Zhao07,Sun-Liang12} to ensure that the phase of source is totally randomized, then our attack is automatically removed. Obviously, this method is the best way for Alice, since it can remove not only our hybrid measurement attack but also other undiscovered attacks based on the random phase of sources, but it may increase the complexity of the system, or introduce other potential and undiscovered loopholes. Note that even an active phase randomization equipment is used by Alice, it is still necessary for her to check the degree of phase randomization in the communication program (but not calibrate it before the communication) to ensure that the phase of source is really randomized in $[0,2\pi)$ and Eve does not break the efficiency of her active phase randomization equipment. The second one is that the legitimate parties carefully design the system parameters to ensure that Eve can not load our attack in these parameter regimes. This method is valid for our hybrid measurement attack, since they know which parameter regimes are secure if they clearly know the parameters of their system, but there may exist other potential hacking strategies so that Eve can also exploit the partially random phase to spy the final key in other parameter regimes. The third one is that the legitimate parties carefully monitor the experimental data but not only estimate the key rate with these experimental data. For example, they can check the rate of gain $Q_\mu/Q_\nu$. In the parameters of Fig.\ref{fig:key_rate}, $Q_\mu/Q_\nu\approx \mu/\nu=4.8$ when Eve is absent, but this rate will be changed to $Q_\mu/Q_\nu\approx7.79$ when Eve is present, which is higher than the expectation 4.8. Furthermore, they also can monitor, with a prior information about the loss of channel, the total gain adn QBER of signal state and decoy state, and so on.

\textbf{Method}

Here we give a simple proof of Eq.\ref{detec}. The state out of Alice can be written as $|\alpha e^{i(\phi+\theta)}/\sqrt{2}\rangle_s\otimes|\alpha e^{i\phi}/\sqrt{2}\rangle_r$, when the two modes pass the BS of Eve (here we simply assume the transmittance of BS is 1/2, in fact Eve can optimize this parameter to maximize her information), the final states are
\begin{equation}
|\frac{1}{2}\alpha e^{i(\phi+\theta)}\rangle_{as}|\frac{1}{2}\alpha e^{i\phi}\rangle_{ar}|\frac{1}{2}\alpha e^{i(\phi+\theta)}\rangle_{bs}|\frac{1}{2}\alpha e^{i\phi}\rangle_{br}.
\end{equation}
If the interferometer of Eve is perfect, the state output of the interferometer can be written as
\begin{equation}
|\frac{1}{2\sqrt{2}}\alpha e^{i\phi}(1+e^{i\theta})\rangle_{D_0}|\frac{1}{2\sqrt{2}}\alpha e^{i\phi}(1-e^{i\theta})\rangle_{D_1}.
\end{equation}
Thus if the SPD of Eve is also perfect, the probability that $D_0$ and $D_1$ click is given by
\begin{equation}
\begin{split}
P_{D_0}&=1-(1-Y_0^E)e^{-\eta_E|\frac{1}{2\sqrt{2}}\alpha e^{i\phi}(1+e^{i\theta})|^2}\\&=1-(1-Y_0^E)e^{-\eta_E|\alpha|^2[1+cos(\theta)]/4},\\
P_{D_1}&=1-(1-Y_0^E)e^{-\eta_E|\frac{1}{2\sqrt{2}}\alpha e^{i\phi}(1-e^{i\theta})|^2}\\&=1-(1-Y_0^E)e^{-\eta_E|\alpha|^2[1-cos(\theta)]/4}.
\end{split}
\end{equation}
Furthermore, for a coherent state $|\alpha\rangle$, the probability distribution of the measured result of homodyne detection can be written as \cite{Sun12}
\begin{equation}
P_x=\sqrt{\frac{2}{\pi\kappa_E^2}}e^{-2[x-\lambda_E|\alpha|cos(\theta)]^2/\kappa_E^2},
\end{equation}
where $\theta$ is the relative phase of signal pulse and local pulse. Thus, it is easy to obtain the third equation of Eq.\ref{detec} for the mode $bs$.

Finally, we list $e^B_{k|j}$ and $e^E_{k|j}$, which are given by
\begin{equation}
\begin{split}
e^{AB}_{k|j}&=[e^B_{kj}]=\begin{bmatrix} 0&1/2&1&1/2\\ 1/2&0&1/2&1\\ 1&1/2&0&1/2\\ 1/2&1&1/2&0\end{bmatrix},\\
e^{AE}_{k|j}&=[e^E_{kj}]=\begin{bmatrix} 0&0&1&1\\ 0&0&1&1\\ 1&1&0&0\\ 1&1&0&0\end{bmatrix}.
\end{split}
\end{equation}

\textbf{Acknowledgements}

This work is supported by the National Natural Science Foundation of China, Grant No. 61072071, and Grant No. 11304391.

\textbf{Author contributions}

S.H.S proposed the main idea of this paper, and does the theoretical analysis and the numerical simulations. M.S.J, X.C.M, C.Y.L and L.M.L contribute the theoretical analysis. All authors agree the contents of the paper.

\textbf{Additional information}

Competing financial interests: The authors declare no competing financial interests.

\end{document}